\documentstyle[sprocl]{article}

\newcommand{\ie}{{\it i.e.}}
\renewcommand{\bar}{\overline}

\newcommand{\eg}{{\it e.g.}}
\newcommand{\MS}{{\rm MS}}

\newcommand{\GeV}{{\rm GeV}}

\def\alh{{\widehat \alpha_s}}
\def\al{\alpha_s}
\def\nf{n_f}
\def\nfh{{\widehat {\nf}}}
\def\npb#1#2#3{{\it Nucl. Phys. }{\bf B #1} (#2) #3}
\def\plb#1#2#3{{\it Phys. Lett. }{\bf B #1} (#2) #3}
\def\prd#1#2#3{{\it Phys. Rev. }{\bf D #1} (#2) #3}
\def\prc#1#2#3{{\it Phys. Reports }{\bf #1} (#2) #3}

\begin{document}

\title{{COMMENSURATE SCALE RELATIONS AND THE ABELIAN
CORRESPONDENCE PRINCIPLE}
 \footnote{\baselineskip=14pt
  Work partially supported by the Department of  Energy, contract
  DE--AC03--76SF00515 and contract DE--FG02--93ER--40762.}}

\author{S. J. BRODSKY}

\address{Stanford Linear Accelerator Center\\
               Stanford University, Stanford, California 94309\\
               E-mail: sjbth@slac.stanford.edu}

\maketitle\abstracts{
Commensurate scale relations are perturbative QCD predictions which relate
observable to observable at fixed relative scales,  independent of the choice of
intermediate renormalization scheme or other theoretical conventions.  A
prominent
example is the ``generalized Crewther relation" which connects the Bjorken and
Gross-Llewellyn Smith deep inelastic scattering sum rules to measurements
of the
$e^+ e^-$ annihilation cross section.  Commensurate scale relations also
provide
an extension of the standard minimal subtraction scheme which is analytic
in the
quark masses, has non-ambiguous scale-setting properties, and inherits the
physical
properties of the effective charge $\alpha_V(Q^2)$ defined from the heavy quark
potential.  I also  discuss a property of perturbation theory, the ``Abelian
correspondence principle", which provides an analytic constraint on
non-Abelian gauge theory
for $N_C \to 0.$}

\section{Introduction}

Quantum Chromodynamics provides an elegant and fundamental description of
hadronic and nuclear interactions in terms of quark and gluon degrees of
freedom.  A common goal of particle and nuclear physics has been to test QCD in
all of its manifestations to as high precision as possible.
A central focus of QCD studies in high energy physics has been the
determination of
the strength of the quark-gluon interaction, as characterized by the
$\alpha_{\overline {\rm MS}}(\mu)$ coupling, defined by convention in a
particular dimensional
regularization scheme.   However, the precision of determining $\alpha_s$
is  limited due to questions of principle in relating physical
measurements to the $\overline {\rm MS}$ coupling.   These problems include
apparent renormalization scale ambiguities,  implementation of finite quark mass
effects,  and the question of the convergence of perturbative expansions
which have
divergent ``renormalon'' $n!$  growth
\cite{tHooft,Mueller,LuOneDim,BenekeBraun}.
Resummations of such divergent series have been proposed, which in turn
highlight the uncertainties
in the behavior of the $\overline{\rm MS}$ coupling at low
momentum scales.   The ambiguities introduced by the scale ambiguities and
scheme conventions of the $\overline {MS}$ scheme are amplified in processes
involving more than one physical scale such as jet observables and
semi-inclusive reactions.
In this talk I will discuss three new theoretical tools which bypass the
above difficulties and have the potential to greatly increase the precision
of QCD tests:  (1) ``commensurate scale relations",  scale-fixed QCD
predictions which
relate observable to observable;  (2) the ``Abelian correspondence
principle", which provides new analytic constraints on QCD predictions; and
(3) the adoption of the effective charge $\alpha_V(Q^2)$ defined from the
heavy quark
potential as a replacement for expansions in
the standard $\overline {\rm MS}$ coupling.   Commensurate scale relations
also provide
an extension of the standard minimal subtraction scheme which is analytic in the
quark masses, has non-ambiguous scale-setting properties, and inherits the
physical
properties of $\alpha_V$.

\section{Commensurate Scale Relations}

Commensurate scale relations relate one physical observables to another physical
observable, and thus must be independent of theoretical conventions such as the
choice of intermediate renormalization scheme.  For example, the ``generalized
Crewther relation", discussed below,  provides a rigorous all-orders
relation between the Bjorken and Gross Llewellyn-Smith sum rules for deep
inelastic scattering  at a given momentum transfer $Q$ to the  annihilation
cross
section $\sigma_{e^+ e^- \to \rm hadrons}(s)$, at a specific ``commensurate"
energy scale \cite{CSR,BGKL}.   The relations between the physical
scales  $Q$ and $\sqrt s$ reflects the fact that the radiative corrections
to the
sum rules and annihilation cross section have different heavy quark
thresholds.
The generalized Crewther relation  can be derived by calculating
the radiative corrections to both the sum rules and $R_{e^+ e^-}$ in the
modified minimal subtraction scheme $\overline{\rm MS}$   and then
algebraically eliminating $\alpha_{\bar MS}(\mu)$.
BLM scale setting is then used to eliminate the $\beta$-dependence of the
coefficients.  However, the relation between
observables at any given order of perturbation theory is independent of the
choice of the choice of renormalization scheme and the initial choice of scale
$\mu$:  obviously,  relations between physical observables cannot depend on
conventions
which theorists choose.  QCD can then be tested in a new fundamental  and
precise way by checking that the observables track both in their relative
normalization and in their commensurate scale dependence.

A helpful tool and notation for relating physical quantities is the effective
charge. Any perturbatively calculable physical quantity can be used to define an
effective charge \cite{Grunberg,DharGupta,GuptaShirkovTarasov} by incorporating
the entire radiative correction into its definition. All effective charges
$\alpha_A(Q)$
satisfy the Gell-Mann-Low renormalization group equation with the same
$\beta_0$
and $\beta_1;$ different schemes or effective charges only differ through
the third
and higher coefficients of the $\beta$ function. Thus, any effective charge
can be
used as a reference running coupling constant in QCD to define the
renormalization
procedure.  More generally, each effective charge or renormalization
scheme, including
$\overline{\rm MS}$, is a special case of the universal coupling function
$\alpha(Q, \beta_n)$
\cite{StueckelbergPeterman,BrodskyLu}.   Peterman and St\"uckelberg have shown
\cite{StueckelbergPeterman} that all effective charges are related to each other
through a set of evolution equations in the scheme parameters $\beta_n.$

For example, consider the Adler function \cite{Adler} for the $e^+ e^-$
annihilation cross section
\begin{equation} D(Q^2)=-12\pi^2 Q^2{d\over dQ^2}\Pi(Q^2),~
\Pi(Q^2) =-{Q^2\over 12\pi^2}\int_{4m_{\pi}^2}^{\infty}{R_{e^+ e^-}(s)ds\over
s(s+Q^2)}.
\end{equation}
The entire radiative correction to this function is defined as
the effective charge
$\alpha_D(Q^2)$ :
\begin{eqnarray}
    D \left( Q^2/ \mu^2, \alpha_{\rm s}(\mu^2/
\Lambda^2_{\overline{\rm MS}}) \right) &=&
D \left (1, \alpha_{\rm s}(Q^2/
\Lambda^2_{\overline {\rm MS}})\right) \label{3} \\
&\equiv&
    3 \sum_f Q_f^2 \left[ 1+ {\alpha_D(Q^2/\Lambda_D^2) \over \pi}
                   \right]
    +( \sum_f Q_f)^2C_{\rm L}(Q^2) \nonumber \\
&\equiv& 3 \sum_f Q_f^2 C_D(Q^2)+( \sum_f Q_f)^2C_{\rm L}(Q^2),
\nonumber
\end{eqnarray}
where $\Lambda_D$ is the scheme-independent effective scale
parameter.  The coefficient $C_{\rm L}(Q^2)$ appears at the third order in
perturbation theory and is related to the ``light-by-light scattering type"
diagrams.  (Hereafter $\alpha_{\rm s}$ will denote the ${\overline{\rm MS}}$
scheme strong coupling constant.)

Similarly, we can define the entire radiative correction to the Bjorken sum rule
as the effective charge $\alpha_{g_1}(Q^2)$ where
$Q$ is the corresponding momentum transfer:
\begin{equation}
\int_0^1 d x \left[ g_1^{ep}(x,Q^2) - g_1^{en}(x,Q^2) \right]
   \equiv {1\over 6} \left|g_A \over g_V \right|
   C_{\rm Bj}(Q^2)
 = {1\over 6} \left|g_A
\over g_V \right| \left[ 1- {\alpha_{g_1}(Q^2/\Lambda_{g_1}^2) \over
   \pi} \right]  .
\end{equation}
It is straightforward to algebraically relate $\alpha_{g_1}(Q^2)$ to
$\alpha_D(Q^2)$ using the known expressions to three loops in the
$\overline{\rm MS}$ scheme. If one chooses the renormalization scale to re-sum
all quark and gluon vacuum polarization corrections into $\alpha_D(Q^2)$,  then
the final result turns out to be remarkably simple:
 $(\widehat\alpha = 3/4\, C_F\ \alpha/\pi = \alpha/\pi)$
\cite{BGKL}
\begin{equation}
\widehat{\alpha}_{g_1}(Q)=\widehat{\alpha}_D(\overline Q^*)-
\widehat{\alpha}_D^2(\overline Q^*)+\widehat{\alpha}_D^3(\overline Q^*) +
\cdots,
\end{equation}
Here
\begin{eqnarray}
\ln \left({ {\overline Q}^{*2} \over Q^2} \right) &=&
{7\over 2}-4\zeta(3)+\left(\frac{\alpha_D (\overline Q^*)}{4\pi}
\right)\Biggl[ \left(
            {11\over 12}+{56\over 3} \zeta(3)-16{\zeta^2(3)}
      \right) \beta_0\cr &&
      +{26\over 9}C_{\rm A}
      -{8\over 3}C_{\rm A}\zeta(3)
      -{145\over 18} C_{\rm F}
      -{184\over 3}C_{\rm F}\zeta(3)
      +80C_{\rm F}\zeta(5)
\Biggr].
\label{EqLogScaleRatio}
\end{eqnarray}
where in QCD $C_{\rm A}=3$, $C_{\rm F}=4/3$.
This relation shows how
the coefficient functions for these two different processes are
related to each other at their respective commensurate scales. The
evaluation of one of them at the appropriate physical scale gives us
information about the second one at the different physical scale.
Notice also that all the $\zeta(3)$ and $\zeta(5)$ dependencies have
been absorbed into the renormalization scale $\overline Q^*$.  We emphasize
that the $\overline{\rm MS}$ renormalization
scheme is used only for calculational convenience; it serves simply as an
intermediary between observables. The renormalization group
property \cite{BogoliubovShirkov,StueckelbergPeterman,GellMannLow} ensures
that the forms of the CSR relations in perturbative QCD are independent
of the choice of an intermediate renormalization scheme.

The Crewther relation was originally derived assuming that the theory is
conformally
invariant; \ie, for zero $\beta$  function. In the physical case, where the
QCD coupling runs,  the non-conformal effects are resummed into the
energy and momentum transfer scales of
the effective couplings $\alpha_R$ and $\alpha_{g1}$.  The coefficients in the
relation between the two effective charges
\begin{equation}
        1- \frac{\alpha_{g_1}(Q)}{\pi} =
\left[ 1+ \frac{\alpha_D(\overline Q^*)}{\pi}\right]^{-1} \ .
\end{equation}
This relation generalizes Crewther's relation to non-conformal
QCD.  Notice that the coefficients which appear in the perturbative
expansion  form a simple
geometric series, and thus do not have a divergent renormalon behavior $n!
\alpha_s^n.$    This is again a special advantage of relating observable to
observable.
The coefficients are independent of
color and are the same in Abelian, non-Abelian, and conformal gauge theory.  The
non-Abelian structure of the theory is reflected in the expression for the scale
$\overline{Q}^{*}$.

The generalized Crewther relation can also be written in the form
\begin{equation} {1\over 3\sum_f Q_f^2} R_{e^+e^-}(s) C_{\rm
Bj}(Q^2)=1+\varepsilon_1(Q^2),
\label{EqBjCrewther}
\end{equation} and
\begin{equation}
 {1\over 3\sum_f Q_f^2}  R_{e^+e^-}(s) C_{\rm GLS}(Q^2)=1+\varepsilon_2(Q^2),
\label{EqGLSCrewther}
\end{equation}
where  $\varepsilon_1$ and $\varepsilon_2$  are small
quantities from NNLO corrections; \eg\ light-by-light scattering contributions.
 The experimental measurements of the
$R$-ratio above the thresholds for the production of $c\overline{c}$-bound
states, together with the theoretical fit performed by Mattingly and Stevenson
\cite{MattinglyStevenson},  provide the empirical constraint
\begin{equation} {1\over 3\sum_f Q_f^2} R_{e^+e^-}(\sqrt s=5.0~{\rm GeV})\simeq
{3\over 10} (3.6\pm 0.1)=1.08\pm 0.03.
\end{equation} and  thus
\begin{equation} {\alpha_{R}^{\rm exp}(\sqrt s=5.0~{\rm GeV})
\over \pi}
 \simeq 0.08\pm 0.03.
\label{Eqalpha}
\end{equation} The prediction for the effective coupling in
the deep inelastic sum rules at the commensurate momentum transfer $Q$
is
\begin{equation} {\alpha_{g_1}^{\rm exp}(Q=12.33\pm 1.20~{\rm GeV})\over \pi}
\simeq {\alpha_{\rm GLS}^{\rm exp}(Q=12.33\pm 1.20~{\rm GeV})\over \pi}
\simeq 0.074\pm 0.026 \ .
\label{EqNumericalBjGLS}
\end{equation}
Measurements of the Gross-Llewellyn Smith sum rule have been carried out
only at relatively small values of $Q^2$
\cite{CCFRL1,CCFRL2}; however, one can use the results of the theoretical
extrapolation \cite{KS} of the experimental data presented in \cite{CCFRQ}:
\begin{equation} {\alpha_{\rm GLS}^{\rm extrapol}(Q=12.25~{\rm GeV})\over
\pi}\simeq 0.093\pm0.042.
\end{equation}
This interval  overlaps with the prediction \cite{Neubert} from  the generalized
Crewther relation.   It is clear that higher precision measurements will be
necessary to fully test these fundamental relations.

Commensurate scale relations allow one to relate any
perturbatively calculable observable, such as the annihilation ratio $R_{e^+
e^-}$, the heavy quark potential and the radiative corrections to structure
function sum rules, to each other without any renormalization scale or scheme
ambiguity \cite{CSR}.    Commensurate scale relations can also be
applied in grand unified theories to make scale-fixed,  scheme invariant
predictions which relate physical observables in different sectors of the
theory.   In each case, commensurate scale
relations connecting the effective charges for observables $A$ and $B$ have the
form $\alpha_A(Q_A) =
\alpha_B(Q_B) \left(1 + r_{A/B} {\alpha_B\over \pi} +\cdots\right),$ where the
coefficient $r_{A/B}$ is independent of the number of flavors $n_F$ contributing
to coupling constant renormalization. The scales of the effective charges that
appear in commensurate scale relations are thus fixed by the requirement that
the couplings sum all of the effects of the non-zero $\beta$ function; the
coefficients in the perturbative expansions in the commensurate scale relations
are thus identical to those of a corresponding conformally-invariant theory with
$\beta=0.$  The method thus has the important advantage of isolating and
``pre-summing" the large and strongly divergent terms in the PQCD  series which
grow as $n! (\beta_0
\alpha_s)^n$, \ie, the infrared renormalons associated with coupling-constant
renormalization \cite{tHooft,Mueller,LuOneDim,BenekeBraun}.  The
renormalization scales $Q^*$ in the BLM method are physical in the sense that
they reflect the mean virtuality of the gluon propagators
\cite{BLM,LepageMackenzie,Neubert,BallBenekeBraun}.
The ratio of scales
$\lambda_{A/B} = Q_A/Q_B$ is unique at leading order and guarantees that the
observables $A$ and $B$ pass through new quark thresholds at the same physical
scale.  One also can show that the commensurate scales satisfy the
transitivity rule
$\lambda_{A/B} = \lambda_{A/C} \lambda_{C/B},$  which ensures that predictions
are independent of the choice of an intermediate renormalization scheme or
observable $C.$

\boldmath
\section{Implementation of  $\alpha_V$ Scheme}
\unboldmath

The physics of commensurate scale relations illuminates the importance of using
an effective charge defined from a physical observable to characterize QCD.  The
central advantage of such a procedure is that predictions which relate one
physical observable to another observable have no ambiguities from theoretical
conventions  such as the choice of renormalization scale or scheme.

The heavy-quark potential $V(Q^2)$ is defined as the
two-particle-irreducible scattering amplitude of test charges; \ie \ the
scattering of two infinitely-heavy quark and antiquark at momentum transfer $t =
-Q^2.$  The relation $V(Q^2) = - 4 \pi C_F
\alpha_V(Q^2)/Q^2$ with $C_F$ given by $C_F=(N_C^2-1)/2 N_C=4/3$ then defines
the effective charge $\alpha_V(Q).$  This coupling can provide a
physically-based alternative to the usual ${\bar {\MS}}$ scheme.
As in the corresponding case of Abelian QED, the scale $Q$ of the coupling
$\alpha_V(Q)$ is identified with the exchanged  momentum. There is thus never
any ambiguity in the interpretation of the scale.   All vacuum polarization
corrections
due to fermion pairs are incorporated in  $\alpha_V$ terms of the usual
vacuum polarization
kernels which are functions of the physical mass thresholds.  An similar
alternative is the effective charge defined from heavy quark
radiation \cite{Uraltsev}.

The-relation of  $\alpha_V(Q^2)$  to the  conventional $\overline {MS}$
coupling  is now known to NNLO \cite{Peter}.
Recently, Gill, Melles, Rathsman and I \cite{BGMR} have derived the required
connection in the form of  a single-scale commensurate scale relation \cite
{Kataev}.
\begin{eqnarray}
\label{eq:csrmsovf}
\alpha_{\overline{\mbox{\tiny MS}}}(Q) & = & \alpha_V(Q^{*}) +
\frac{2}{3}N_C{\alpha_V^2(Q^{*}) \over \pi}
\nonumber \\
&& +\Bigg\{ -\left(\frac{5}{144}+\frac{24\pi^2-\pi^4}{64}-
       \frac{11}{4}\zeta_3\right)N_C^2\nonumber \\
&& +\left(\frac{385}{192}-\frac{11}{4}\zeta_3\right)C_F N_C\Bigg\}
 {\alpha_V^3(Q^{*}) \over \pi^2} \nonumber \\ & = &
\alpha_V(Q^{*}) + 2{\alpha_V^2(Q^{*}) \over \pi} + 4.625 {\alpha_V^3(Q^{*})
\over \pi^2} ,
\end{eqnarray} above or below the quark mass threshold.  The coefficients in the
perturbation expansion have their conformal values, \ie, the same coefficients
would occur even if the theory had been conformally invariant with
$\beta=0$ and thus do not contain the diverging
$(\beta_0\alpha_{\mbox{\scriptsize{s}}})^n n!$  growth characteristic of an
infrared renormalon \cite{Kataev}.  The next-to leading order (NLO) coefficient
$\frac{2}{3}N_C$ is a feature of the non-Abelian couplings of QCD and is not
present in QED.   Here
\begin{eqnarray} Q^* & = & Q\exp\left[\frac{5}{6}  +
    [(\frac{35}{32}-\frac{3}{2}\zeta_3)C_F -
        (\frac {19}{48} -\frac{7}{4}\zeta_3) N_C]
               \frac{\alpha_V}{\pi} + \cdots
             \right]
\end{eqnarray}
For $N_C= 3$ we have $\ln Q^*/Q =   5/6 + 4.178 \alpha_V/\pi.$
The factor $e^{5/6} \simeq 0.4346$ is the ratio of commensurate scales between
the two schemes to leading order.   It arises because of the convention used in
defining the modified minimal subtraction scheme. The scale in the $\bar {\MS}$
scheme is thus a factor
$\sim 0.4$ smaller than the physical scale. The coefficient $2 N_C/3$ in the NLO
coefficient is a feature of the non-Abelian couplings of QCD; the same
coefficient occurs even if the theory were conformally invariant with
$\beta_0=0.$

Using the above QCD results, we can transform any NNLO  prediction
given in $\overline{MS}$ scheme as a scale-fixed expansion in
$\alpha_V(Q)$~\cite{BEGKS}.
We can derive the connection between the $\overline{MS}$ and $\alpha_V$
schemes for  Abelian perturbation theory using  the limit $N_C \to 0$ with
$C_F\alpha_s$ and $N_F/C_F$ held fixed~\cite{BrodskyHuet} (see Section 4).
In this case\begin{eqnarray}
\widehat \alpha_{\overline{\mbox{\tiny MS}}}(Q) & = & \widehat \alpha_V(Q^{*})
\end{eqnarray} with
\begin{eqnarray} Q^* & = & Q\exp\left[\frac{5}{6}  +
    [(\frac{35}{32}-\frac{3}{2}\zeta_3)
                       \frac{\widehat\alpha_V}{\pi}+ \cdots
             \right] \ .
\end{eqnarray}

The use of $\alpha_V$  as the expansion parameter with BLM scale-fixing has been
found to be  valuable in lattice gauge theory, greatly increasing the
convergence of perturbative expansions relative to those using the bare lattice
coupling \cite{LepageMackenzie}.  Recent lattice calculations of the
$\Upsilon$- spectrum \cite{Davies} have been used to determine the normalization
of the static heavy quark potential and its effective charge $
\alpha_V^{(3)}(8.2 \GeV) = 0.196(3)$ where the effective number of light
flavors is
$n_f = 3$.  A recent determination \cite{Davies} of the corresponding
modified minimal subtraction coupling evolved to the
$Z$ mass is given by $ \alpha_{\bar{\MS}}^{(5)}(M_Z) = 0.1174(24)$.

Thus a high precision value for $\alpha_V(Q^2)$  at a specific scale is
available.  Predictions for other QCD observables can be  directly
referenced to this value,  without the scale or scheme ambiguities, greatly
increasing the precision of QCD tests. We can anticipate that  eventually
nonperturbative
methods such as lattice gauge theory or discretized light-cone quantization will
provide a complete form for the heavy quark potential in $QCD$.   It
is reasonable to assume that $\alpha_V(Q)$ will not diverge at small space-like
momenta. One possibility is that $\alpha_V$ stays relatively constant
$\alpha_V(Q) \simeq 0.4$ at low momenta, consistent with  fixed-point behavior.
There is, in fact, empirical evidence for freezing of the $\alpha_V$  coupling
from the observed systematic dimensional scaling behavior of exclusive
reactions \cite{BJPR}.   If this is in fact the case, then the range of QCD
predictions can be extended to quite low momentum scales, a regime normally
avoided because of the apparent singular structure of perturbative
extrapolations.

There are other advantages of the $V$-scheme:
\begin{enumerate}
\item
Perturbative expansions in $\alpha_V(Q^*)$ cannot have any
$\beta$-function dependence in their coefficients  since all vacuum polarization
contributions to the running are already summed into the definition of the
potential and  the effective coupling.   There is thus never any scale
ambiguities.   The  value of the scale  $Q^*$ reflects the mean virtuality  of
the exchanged gluons  in the Feynman amplitude.   Since coefficients involving
$\beta_0$ cannot occur in an expansions in $\alpha_V$,  diverging infrared
renormalons of the form $\alpha^n_V\beta_0^n n!$  cannot occur.   The
general convergence properties of the scale $Q^*$ as an expansion in $\alpha_V$
is not known \cite{Mueller}.

\item
The effective coupling $\alpha_V(Q^2)$ incorporates vacuum polarization
contributions with finite fermion masses.  When continued to timelike
momenta, the coupling has the correct analytic dependence dictated by particle
production in the $t$ channel.  Thus since  $\alpha_V$ incorporates  quark mass
effects exactly, it avoids the problem of explicitly computing and resuming
quark mass corrections.

\item Eq. (\ref{eq:csrmsovf}) is technically only valid far above and below
a heavy quark threshold.  However, the same equation can be used to
define an  analytically-extended $\overline {MS}$  scheme at any scale $Q$.
The new
modified scheme inherits all of the good properties of the $\alpha_V$ scheme,
including its correct analytic properties as a function of the quark masses and
unambiguous scale fixing \cite{BGMR}.

\item
The use  of $\alpha_V$ at any stage allows a simple connection to the
Abelian theory via the $N_C \to 0$ limit.   I discuss this further in the next
section.

\item
Computations in different sectors of the Standard Model have been
traditionally carried out using different renormalization schemes.   The
traditional QED scheme is equivalent to $\alpha_V$.  However, in a grand
unified theory, the forces between all of the particles in the fundamental
representation should become universal above the grand unification scale.
Thus it is
natural to use $\alpha_V$ as the effective charge for all sectors of a grand
unified theory since unification should occur in $\alpha_V(Q^2)$  rather than in
a convention-dependent coupling such as $\alpha_{\overline {MS}}$.

\item
The  $\alpha_V$  coupling is the natural expansion parameter for
processes involving non-relativistic momenta, such as heavy quark production at
threshold where the Coulomb interactions, which are enhanced at low relative
velocity $v$ as $\pi \alpha_V/v$, need to be re-summed.
\cite{Voloshin,Hoang,Fadin}
The threshold corrections
to heavy quark production in $e^+ e^-$ annihilation depend directly on
$\alpha_V$ at specific scales $Q^*$.  Two distinct ranges of scales arise
as arguments of
$\alpha_V$ near threshold: the relative momentum of the quarks governing the
soft gluon exchange responsible for the Coulomb potential, and a high momentum
scale approximately equal to twice the quark mass for the corrections induced by
hard gluon exchange \cite{Hoang}.  One thus can use threshold production to
obtain a direct
determination of $\alpha_V$ even at low scales.  The corresponding QED results
for $\tau$ pair production allow for a measurement of the magnetic moment of the
$\tau$ and could be tested at a future $\tau$-charm
factory \cite{Voloshin,Hoang}.

\item
The effective NRQCD Hamiltonian  is  effectively written in $\alpha_V$
scheme.

\end{enumerate}

One can also apply commensurate scale relations in $\alpha_V$ to the domain of
exclusive processes at large momentum transfer such as the  form factors and
the photon-to-pion transition form factor at
large momentum transfer \cite{BLM,BJPR}  and exclusive weak decays of heavy
hadrons in QCD \cite{Henley}.   Each gluon
propagator with four-momentum $k^\mu$ in the hard-scattering quark-gluon
scattering amplitude is associated with the coupling $\alpha_V(k^2)$ since the
gluon  exchange propagators closely resembles the interactions encoded in the
effective potential $V(Q^2)$. [In Abelian theory this is exact.]
Commensurate scale
relations can then be
established which connect the hard-scattering subprocess amplitudes which
control exclusive processes to other QCD observables.

\section{QCD in the Limit of Small Number of Colors.}

A remarkable property of perturbative QCD, first demonstrated by  't Hooft
\cite{tHooft} is
that the theory is dominated by diagrams with planar topology in the limit $N_C
\to \infty$.  In this limit, the dynamics  of the theory is effectively
constrained by the
color degrees of freedom.  Recently Patrick Huet and I \cite{BrodskyHuet}
have explored
the general properties of perturbative QCD expressions taken as analytic
functions of  $z = N^2_C$.   We found several unexpected features  of the
$SU(N_C)$ theory which provide useful constraints on non-Abelian gauge theory,
including an interesting Abelian limit for $N_C^2=0$

It is useful to introduce rescaled couplings and flavor number
$\widehat \alpha_s = C_F \alpha_s,$ \break $\widehat n_f = Tn_F/C_F,$   where
$C_F={N_C^2-1\over 2 N_C}$  is the fundamental Casimir constant and  $T=1/2$.
At large $N_C^2$, $\widehat \alpha$ incorporates the rescaling of the
coupling advocated by  't Hooft.

The expansion of QED predictions for color-averaged quantities  in the rescaled
coupling have the form
\begin{equation}
C_{n,\ell} = {\widehat \alpha_s^n \widehat n_f^\ell \over
(N_C^2-1)^{\ell-n}}
\Sigma_{i=1}^{2^n}(-1)^{n-e_i}(N_C^2)^{\widetilde \omega_i},
\end{equation}
where $\widetilde \omega_i$  is an index computed from the topology of the
component color graph which is obtained by replacing the gluons by $ e_i$
``double lines" using the Cvitanovic-Mandula rules. The maximum value for the
index occurs when all gluons are replaced  with double $q\bar q$ lines as
in a $U(N)$
theory.  For planar graphs, $C_{n,\ell}$ grows maximally at large $N_C$  as
${\widehat \alpha_s}^n  {\widehat n_f}^\ell N_C^2$   which is  the `t Hooft
limit.  On the other hand, for
$N_C^2 \to 0$,  the component diagrams which dominate the color factor have
$\widetilde \omega= 0$  and occur only from color graphs which have a
``tree structure.''
Thus for $N_C^2 \to 0,$   the coefficient of ${\widehat \alpha_s}^n  {\widehat
n_f}^\ell$  is a finite constant and is identical to the coefficients of an
Abelian
theory.  The two limits essentially bound the
polynomial behavior of perturbation theory for large and small color.  The only
analytic singularity occurs at $N_C^2 \to 1$ where $SU(N_C)$  becomes
undefined.

The $N_C^2 \to 0$ limit reduces the non-Abelian theory to an Abelian theory
dominated by the coupling of the $N_C - 1$  diagonal gluons of the adjoint
representation. The small-$N_C$  limit of $SU(N_C)$ reflects the  coupling
of $N_C-1$.  Abelian gluons and thus has the group structure
$\lim_{N_C \to 0} [U(1]^{N_C-1} \sim U(1)^{-1}$, \ie: $-1$ Abelian gluons.
The  $N_C \to 0$ theory resembles QED; however,  in high order
graphs involving fermion loops, there is an ``offset"  factor relative to
the QED
value calculated with ${\widehat \alpha_s} \to \alpha_{QED}$ and ${\widehat n_f
}\to n_{\rm leptons}$ in QED.  The offset factor is easily evaluated by counting
the number of tree diagrams contributing to the color weight.  For example,  a
color diagram originating from  a ring of $\ell=3$ fermion loops interconnected
pair-wise by $p,q,$  and $ r$ gluons has the offset factor
$(p-1) (q-1) (r-1) -pqr$.

As an example, consider the QCD prediction for the ratio of the annihilation
cross section to the  point-like limit
$R_{e^+ e^-}$  in  the ${\overline {MS}}$ scheme.  In terms of $\widehat
\alpha_s$  and $\widehat n_f$,
\begin{equation}
  R_{e^+ e^-}(Q^2)/(N^2_C-1) =
 \sum_I^{\nfh} Q^2_I \lbrace 1 +{\alh(Q)\over \pi} \widehat F_2
 +{\alh^2(Q)\over \pi^2}\widehat F_3 + \cdots \rbrace + \cdots
 \end{equation}
with $\widehat F_n = F_n/C_F^{n-1}$. Specifically,
\begin{equation}
\widehat F_2 = {3\over 4}\ \ {\rm and} \ \ \ \widehat F_3 = -{3
\over 32}+ \left({123 \over 32} - {11\over 4}
\zeta(3)\right)({N_C^2\over N_C^2-1}) +
\left( {-11\over 8} + \zeta(3)\right)\nfh\, .
\end{equation} For $N_C^2 \to 0$, these forms coincide with the QED coefficients
$F_2^{QED}$  and $F_3^{QED}$ with $\alpha_{QED}= \widehat\alpha_s$ and $n_{\rm
leptons} = \widehat n_f$.   The coefficients of $\al^3$ in the expansion above
has been computed and the corresponding $\widehat F_4$ also coincides with its
QED counterpart \cite{ss}.  In the next order where the Casimir
$d_{abc}^2$ appears, the
QED result is $1/2$  of the  $N_C^2 \to 0$ limit of the QCD production due
to the
offset factor.

The simple structure of the color coefficients in the rescaled quantities
$\widehat\alpha_s$ and  $\widehat n_f$ provides a constraint on Pad\'e and other
methods which resum perturbation theory since no coefficient can grow faster
then $N_C^2$.

\section{The Abelian Correspondence Principle}\

The non-trivial analytical limit of perturbative QCD expressions at small number
of colors provides a new type of ``correspondence principle":  QCD predictions
must coincide analytically with predictions of the corresponding Abelian theory
at $N_C \to 0. $  In addition to providing a boundary condition and useful
check on non-Abelian analyses,  there are a number of important physical
implications:

\begin{enumerate}
\item
Perturbative  QCD results, such as factorization theorems for hard
inclusive and exclusive reactions, evolution equations,   and  results
derived from the operator
product expansion  are immediately applicable to QED.  Similarly,
physical principles controlling the high energy interactions of hadrons in
QCD such as,
diffraction,  hard pomeron and odderon exchange,  color transparency and
intrinsic heavy particle Fock states  all have physical analogs for the
interactions of neutral atoms in QED.  Conversely, phenomena in QED atoms such
as van der Waals interactions, co-mover coalescence, and  the Lamb shift,
predict analogous phenomena in QCD.

\item
 The treatment of renormalization  schemes and scales in perturbative QCD
must match those of QED at $N_C\to 0.$ In QED it is traditional to define the
fundamental effective charge of the theory $\alpha_{QED}(Q^2)$ as the coupling
which appears in the potential between two massive test charges:  $V(Q^2) =  -
{4 \pi Z_1 Z_2 \alpha_{QED}(Q^2)\over Q^2}$ where $Q^2 = -q^2$ is the space-like
momentum transfer squared. and normalize it to the  measured value at
$Q^2 =0\,:\,
\left[\alpha_{WED}(0)\right]^{-1}= 137.0359895 (61)$ \cite{PDF}.   In the
QED scheme,
all vacuum polarization effects which normalize the photon propagator are summed
into $\Pi(Q^2)$.  There is thus no scale ambiguity and fermion pair masses are
treated exactly.   As we have seen in Section 3, these constraints are fulfilled
when $\alpha_V$ is used as the effective charge in QCD: perturbative QCD
expressions in the $\alpha_V$ scheme have the correct Abelian
correspondence limit with
QED expressions in the $\alpha_{QED}$  scheme.

\end{enumerate}

The above analyses of the color weights of $SU(N_C) $ gauge theory and the
Abelian limit at $N_C \to 0$ apply to any order in perturbation theory.   The
coefficients in $\widehat\alpha$ and $\widehat n$ which are finite in $N_C$
are given by the Abelian theory.  Alternatively, we
can use the general $N_C$ analysis to expand QCD expressions at small $N_C$,
starting with the QED prediction as the initial approximation. The most
interesting  questions center on whether the simple analytic properties of
perturbation theory also hold
for nonperturbative QCD predictions, such as those calculated from
instanton effects.
More generally, does confinement or QCD phase transitions lead to non-analytic
behavior in $N_C^2$ not present in all-order perturbative analyses?

\section*{Acknowledgments}

The work on the generalized Crewther relation reported in
Section 2 is based on a collaboration with H.-J. Lu, A. Kataev
and G.~Gabadadze.   The results in Section 3 are based on collaborations with
Mandeep Gill, Michael Melles, Chueng Ji, Alex Pang, and Dave Robertson.  The
work on QCD at small number of colors reported in Section 4 is based on work
with Patrick  Huet and Nicolaos Toumbas.  I also thank N. Toumbas, J.~Paleaz,
E.~Sather, M.~Beneke, V.~Braun,  A.~Hoang,  H.~K\"uhn,  T.~Tuebner,
G.~P.~ Lepage, G.~Mirabelli, A.~Mueller, D.~M\"uller, O.~Puzyrko,  and
W.-K.~Wong for helpful discussions. This work is supported in part by the
Department
of Energy, contract DE--AC03--76SF00515.

\section*{References}

\begin{sloppy}
\begin{raggedright}

\end{raggedright}
\end{sloppy}
\end{document}